\documentclass[11pt,a4paper]{article}

\usepackage[margin=2.3cm]{geometry}
\usepackage{graphicx}
\usepackage{amsmath,amssymb,bm}
\usepackage{upgreek}
\usepackage{mathrsfs}
\usepackage{mathcomp}
\usepackage[compress]{cite}
\usepackage[hidelinks]{hyperref}

\title{\bfseries Increase of Magnetic Trap Loading Efficiency for $\bm{^{39}\mathrm{K}}$ Bose--Einstein Condensation Experiments}

\author{%
Yihao Cheng$^{1,2,\dagger}$, Xiaoyang Shen$^{\dagger}$, Weixin Li$^{1,2}$, Hao Du$^{1,2}$,\\Hongwu Zhao$^{1,2,3}$, and Lin Xia$^{1,3,*}$\\[0.8em]
\small $^{1}$Beijing National Laboratory for Condensed Matter Physics, Institute of Physics,\\\small Chinese Academy of Sciences, Beijing 100190, China\\
\small $^{2}$University of Chinese Academy of Sciences, Beijing 100049, China\\\vspace{0.5em}
\small $^{3}$Songshan Lake Materials Laboratory, Dongguan 523808, China\\
\small $^{\dagger}$These authors contributed equally to this work.\\
\small $^{*}$Corresponding author: linxia@iphy.ac.cn
}
\date{}

\begin{document}
\maketitle

\begin{abstract}
During the experimental sequence, a large atom number is important for the creation of high-quality Bose--Einstein condensates. We report an experimental increase of about 15\% in magnetic-trap (MT) loading efficiency by optimizing the MT loading process. The loading time is much shorter than the lifetime of the atoms in the MT, which means atom loss during loading due to the finite trap lifetime can be neglected. We obtain an almost pure $^{39}\mathrm{K}$ condensate with $1.2\times10^{5}$ atoms after evaporative cooling in our optical trap.
\end{abstract}

In recent years, ultracold $^{39}\mathrm{K}$ atomic gases\cite{roati2007bose} have been widely used to study various quantum phenomena. In 2018, Cabrera \textit{et al.} observed quantum liquid droplets by combining two $^{39}\mathrm{K}$ Bose--Einstein condensates (BEC) in different states\cite{cabrera2018quantum}. In the same year, Semeghini \textit{et al.} observed self-bound $^{39}\mathrm{K}$ quantum droplets in free space and showed the importance of quantum fluctuations in the stabilization of droplets\cite{semeghini2018self}. In 2024, Yu \textit{et al.} realized a two-dimensional Bose glass in a quasicrystalline optical lattice and observed its transition to a superfluid\cite{yu2024observing}. Other active research directions including Bose polarons\cite{morgen2025quantum}, few-body physics\cite{morgen2025three} and strongly interacting many-body quantum systems\cite{konstantinou2026suppression,morris2025scaling} have also been studied. The widely tunable interaction of $^{39}\mathrm{K}$ makes it an ideal platform for exploring quantum many-body physics\cite{derrico2007feshbach}. The stable production of high-quality BEC is essential for further research.

The realization of BEC requires the preparation of a sufficiently large number of atoms at low temperature and high phase-space density. There are two main reasons. First, for a three-dimensional harmonic-oscillator potential, the transition temperature is given by $k_BT_c \approx 0.94\hbar\bar{\omega}N^{1/3}$\cite{pethick2002bose}, where $\bar{\omega}$ is the mean frequency and $N$ is the atom number. $T_c$ rises as $N$ increases, so it will be easier to reach the transition point with more atoms. Second, a conventional cooling technique to obtain BEC is forced evaporative cooling\cite{ketterle1996evaporative}, which is usually realized by removing the high-energy atoms from the trap\cite{anderson1995observation,pereira2001bose,landini2012direct}. A huge fraction of atoms is lost during this stage, so a large initial atom number is necessary. In a typical BEC experiment, atoms are first cooled by laser cooling and then transferred into a conservative trap for further cooling\cite{davis1995bose,bradley1995evidence}. Magnetic trap (MT) provides an effective way of confining cold neutral atoms\cite{migdall1985first}. The efficiency of loading laser-cooled atoms into the MT plays an important role in the preparation of BEC because it determines the initial atom number and temperature for subsequent cooling procedures.

In this work, we report the realization of a $^{39}\mathrm{K}$ Bose--Einstein condensate. We focus on the MT loading process and improve the efficiency by applying a higher magnetic field gradient. This optimized sequence provides a larger atomic sample for optical dipole trap loading and evaporative cooling.

Our cooling and repumping beams are provided by TOPTICA TA pro, whose output power is $2.3\,\mathrm{W}$. The main beam is locked to $^{39}\mathrm{K}$ transition line $4S_{1/2}, F=2 \to 4P_{3/2}, F'=3$ by modulation transfer spectroscopy (MTS)\cite{shirley1982modulation} and is split into several beams. The beams are later frequency shifted by acousto-optic modulators (AOM). After passing those AOMs, the beams are combined and delivered to the vaccum chamber. In addition, a low-power beam is split from the main laser beam for absorption imaging.

We use a pair of anti-Helmholtz coils to create a quadrupole MT for atom confinement. The trapping mechanism originates from the Zeeman interaction between the atomic magnetic moment and the non-uniform magnetic field. For atoms in low-field seeking states, the magnetic potential can be written as $U(\mathbf{r}) = \mu_{\mathrm{eff}} |\mathbf{B}(\mathbf{r})|$, where $\mu_{\mathrm{eff}}$ is the effective magnetic moment of the trapped state and $\mathbf{B}(\mathbf{r})$ is the position-dependent magnetic field. The field vanishes at the trap center and increases approximately linearly with the distance from the center. Therefore, the low-field seekers experience a restoring force toward the local minimum and can be trapped at the center. The magnetic field gradient is controllable by changing the current applied to the coils. Our MT provides a magnetic-field gradient of $1.23\,\mathrm{G\cdot cm^{-1}\cdot A^{-1}}$ along the vertical, i.e. axial, direction, which is sufficient to provide effective confinement.

As the beginning of the experimental sequence, the $^{39}\mathrm{K}$ atoms are first collected in the magneto-optical trap (MOT). Then the atomic cloud is compressed by increasing the magnetic field gradient and adjusting the laser intensity and detuning, in order to improve spatial density. After that, gray molasses (GM)\cite{salomon2013gray} is applied to achieve a lower temperature. About $1\times10^9$ atoms are obtained after the GM stage. Subsequently, the atoms are optically pumped to the $\lvert F=1,m_F=-1\rangle$ state in preparation for loading into the MT.

The MT current is driven by a programmable DC power supply. We use an analog input signal to control the output current and use an IGBT switch in the circuit to rapidly switch the current on and off. After optical pumping, all lasers are turned off, the input signal rises and the IGBT switch is turned on. The output current rises to $26\,\mathrm{A}$ in $1\,\mathrm{ms}$, corresponding to an axial magnetic field gradient of $32\,\mathrm{G\cdot cm^{-1}}$. In this process, most of the atoms are captured while the remaining atoms fall out of the trapping region. The trap is first held for a short time to allow untrapped atoms to escape and the cloud to rethermalize. In the end, the current is suddenly switched off, and an absorption image is taken after $7\,\mathrm{ms}$ of ballistic expansion to calculate atom numbers.

\begin{figure}[htbp]
    \centering
    \includegraphics[width=0.55\linewidth]{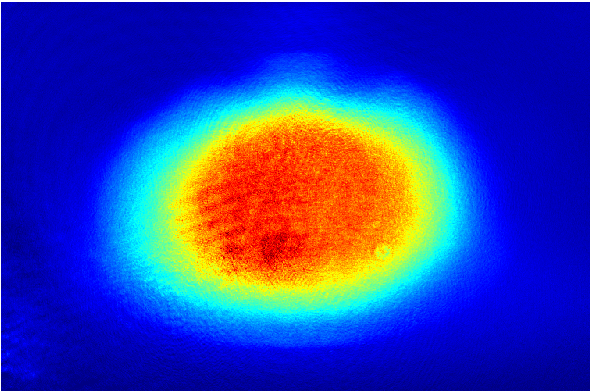}
    \caption{Absorption image of the $^{39}\mathrm{K}$ atomic cloud after MT loading. The picture is taken after $7\,\mathrm{ms}$ free expansion. The picture size is $11.84\times7.70\,\mathrm{mm^2}$.}
    \label{fig:1}
\end{figure}

Fig.~\ref{fig:1} shows an absorption image of the atomic cloud after MT loading. At a coil current of $26\,\mathrm{A}$, we get $7.95\times10^8$ atoms loaded into the MT. There are two main reasons for the atom loss. One is that the atomic cloud has a high optical depth after the GM stage. As a result, the pumping light is completely absorbed before reaching the central part of the cloud, which limits the optical pumping efficiency. The other is that the magnetic field gradient is not sufficient. The calculated gradient is ideal only near the trap center, where the field remains linear. In regions close to the edge of the trap, the atoms experience a lower effective gradient and are more likely to escape.

\begin{figure}[htbp]
    \centering
    \includegraphics[width=0.6\linewidth]{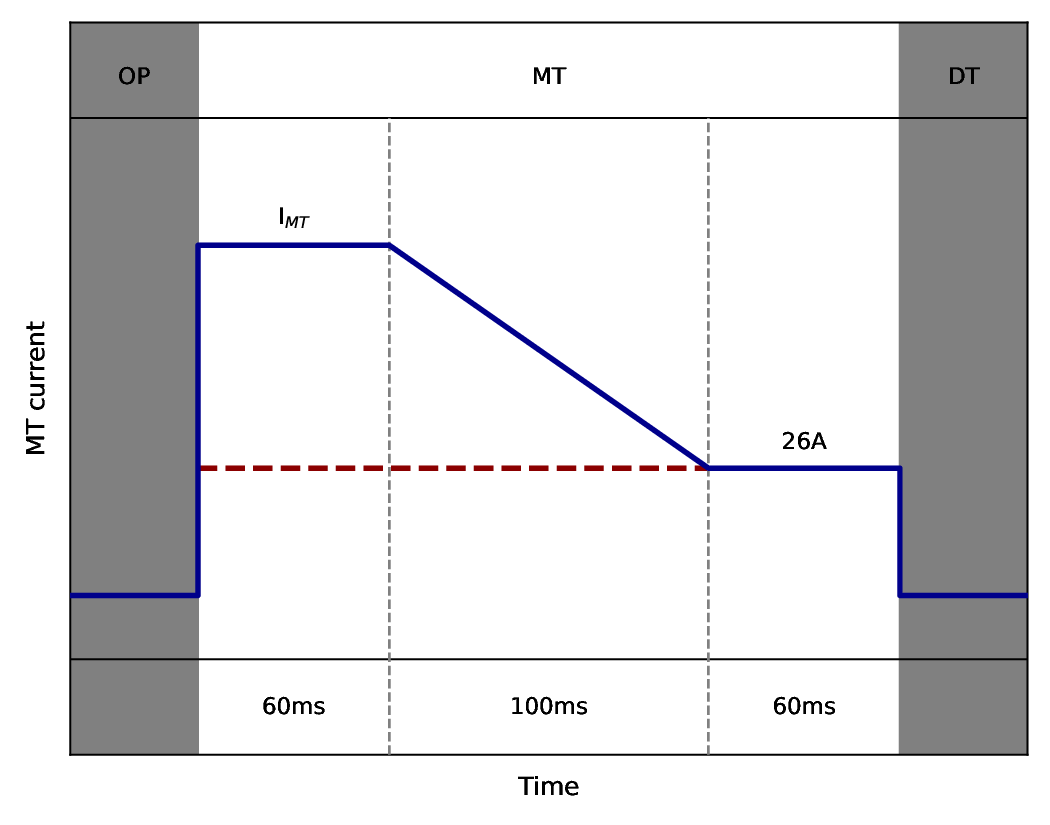}
    \caption{Sequence of the MT loading process. Red dashed line: initial sequence with the current fixed at $26\,\mathrm{A}$. Solid line: optimized sequence, in which the current is first held at a higher value $I_{MT}$ for $60\,\mathrm{ms}$, then linearly ramped down to $26\,\mathrm{A}$ in $100\,\mathrm{ms}$, and held for another $60\,\mathrm{ms}$.}
    \label{fig:2}
\end{figure}

We optimize the MT loading process by increasing the coil current. Fig.~\ref{fig:2} shows the optimized sequence. The loading process consists of three steps. In the first $60\,\mathrm{ms}$, current is held at a value $I_{MT}$ larger than $26\,\mathrm{A}$ to create a higher magnetic field gradient. During the following $100\,\mathrm{ms}$, the current is linearly ramped down to $26\,\mathrm{A}$, ensuring that atoms in other Zeeman states are not trapped, which is necessary for making a pure atomic sample. In the last step, the current is held at $26\,\mathrm{A}$ for another $60\,\mathrm{ms}$, allowing atoms in undesired states to fall out.

\begin{figure}[htbp]
    \centering
    \includegraphics[width=0.68\linewidth]{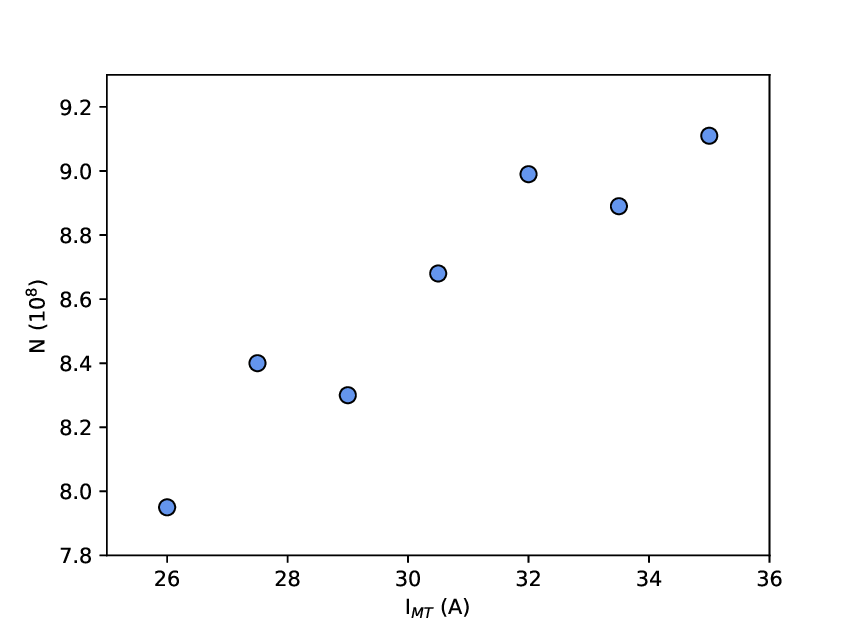}
    \caption{Atoms loaded into the MT with different $I_{MT}$. By increasing $I_{MT}$ from $26\,\mathrm{A}$ to $35\,\mathrm{A}$, the number of loaded atoms improves by 15\%. Three measurements are taken for each point.}
    \label{fig:3}
\end{figure}

For different $I_{MT}$ values, we measure the atom number at the end of the MT loading process. Measurements are taken at $I_{MT} = 26, 27.5, 29, 30.5, 32, 33.5, 35\,\mathrm{A}$, as shown in Fig.~\ref{fig:3}. Initially, only $7.95\times10^8$ atoms are loaded into the trap. By applying the optimized sequence and increasing $I_{MT}$ to $35\,\mathrm{A}$, the number of the trapped atoms reaches $9.11\times10^8$. Compared with the initial loading process, the atom number is increased by about 15\%. The increase mainly results from the enhanced magnetic field gradient at higher $I_{MT}$, which creates a deeper trapping potential, making it more difficult for atoms, especially the ones near the edge of the trap, to escape and thereby increasing the number of atoms captured. Meanwhile, only slight heating is observed ($15\,\mathrm{\mu K}$ at $26\,\mathrm{A}$, $20\,\mathrm{\mu K}$ at $35\,\mathrm{A}$).

\begin{figure}[htbp]
    \centering
    \includegraphics[width=0.68\linewidth]{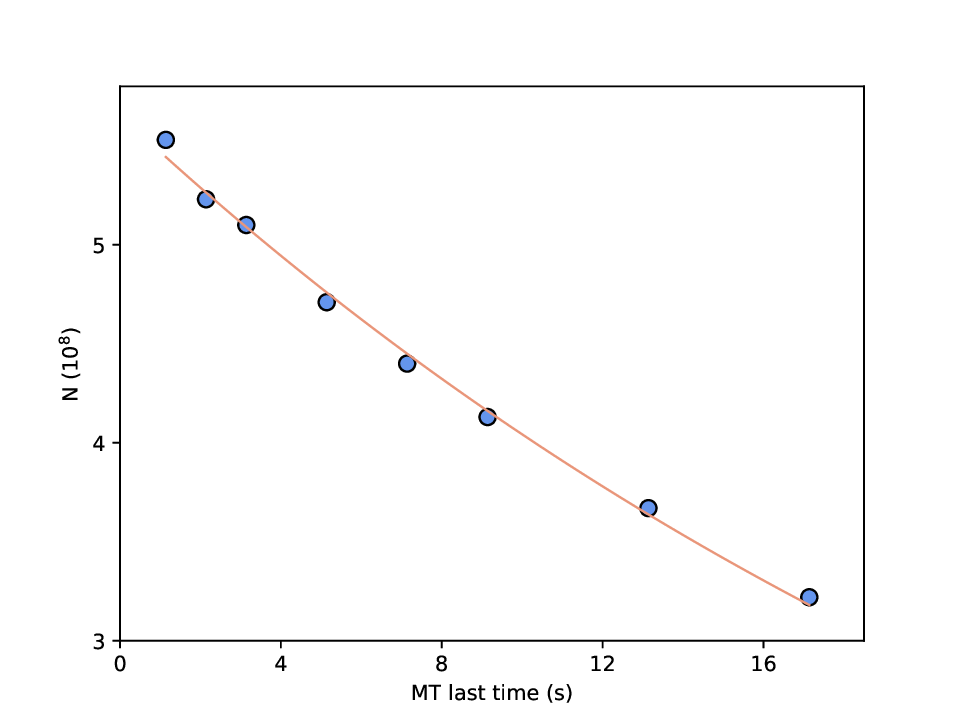}
    \caption{Atom number decay in the MT at $26\,\mathrm{A}$. The solid line is an exponential fit, yielding a $1/e$ lifetime of $29.8\,\mathrm{s}$.}
    \label{fig:4}
\end{figure}

We also characterize the lifetime of the atoms in the MT at a current of $26\,\mathrm{A}$ by measuring the atom number after different MT holding times. Results are shown in Fig.~\ref{fig:4}. The atom number decreases with increasing MT holding time. An exponential fit gives a $1/e$ lifetime of $29.8\,\mathrm{s}$, which is much longer than the MT loading time. Therefore, atom loss due to the finite trap lifetime in the loading process can be neglected.

In addition to improving the MT loading efficiency, this optimized sequence can also serve as a method of diagnosing the optical pumping efficiency. When the optical pumping efficiency remains low after excluding possible imperfections in the light polarization and in the direction, magnitude and homogeneity of the magnetic field, this sequence can be used to diagnose. A higher magnetic field gradient is recommended to obtain a stronger signal. If the number of atoms loaded into the MT increases significantly, it indicates that a considerable amount of atoms escape from the trap under the original experimental parameters. In this case, the actual optical pumping efficiency can be measured more reliably.

\begin{figure}[htbp]
    \centering
    \includegraphics[width=0.90\linewidth]{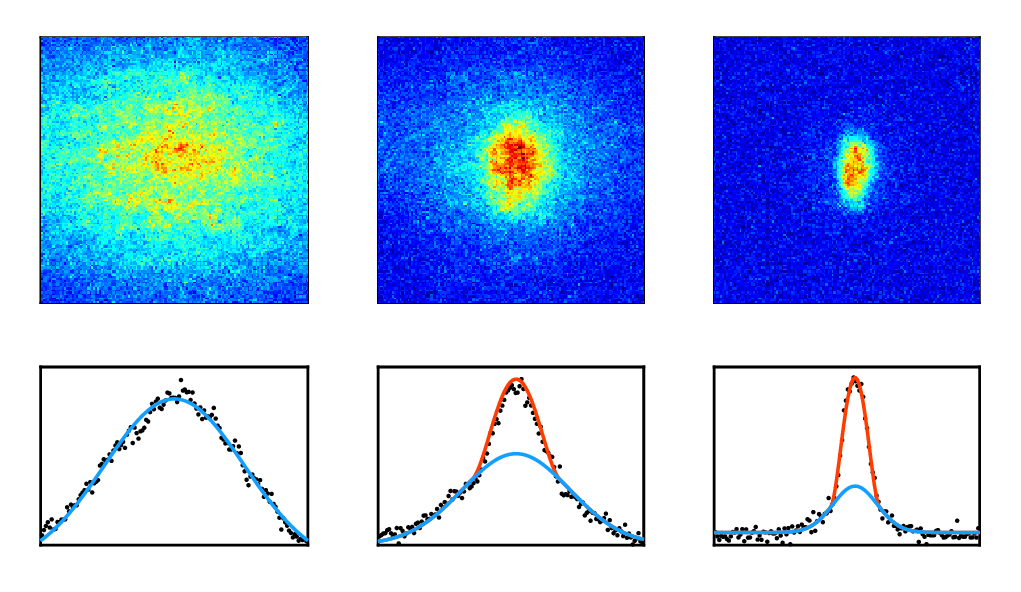}
    \caption{Absorption images showing the BEC transition process during the evaporative cooling stage. Blue curves are thermal fits, and red curves are Thomas--Fermi fits. From left to right, the cloud evolves from a thermal gas to an almost pure condensate. Pictures are taken after $14\,\mathrm{ms}$ free expansion. The picture size is $1.14\times1.14\,\mathrm{mm^2}$.}
    \label{fig:5}
\end{figure}

After the MT stage, the atoms are transferred into a $1064\,\mathrm{nm}$ optical dipole trap with about $20\,\mathrm{W}$ laser power. During the evaporative cooling process, Feshbach field is on. In $1\,\mathrm{s}$, the power of the optical trap is reduced, and an almost pure $^{39}\mathrm{K}$ condensate with $1.2\times10^{5}$ atoms is obtained, as shown in Fig.~\ref{fig:5}.

In conclusion, we realize a $^{39}\mathrm{K}$ Bose--Einstein condensate, with particular emphasis on optimizing the MT loading efficiency. By applying an optimized sequence with a higher MT current, the number of atoms loaded into the trap increases by 15\%. The lifetime of atoms in the MT is much longer than the loading time, which means atom loss arising from the finite trap lifetime can be ignored. The optimized sequence increases the initial number of atoms for dipole trap loading, providing a better condition for evaporative cooling, and thereby contributing to the efficient production of BEC. This work provides a good foundation for our future investigations of quantum phenomena in ultracold $^{39}\mathrm{K}$ gases.

\section*{Acknowledgments}
\begingroup
\emergencystretch=2em
This work is supported by the National Key Research and Development Program of China (2021YFA\allowbreak1400900 and 2021YFA0718302).
\endgroup

\end{document}